\documentclass{ws-procs975x65}
\newcommand{\CD}{{\cal D}}
\newcommand{\CR}{{\cal R}}
\newcommand{\average}[1]{\left\langle #1 \right\rangle_\CD}

\newcommand{\initial}[1]{{#1_{\rm \bf i}}}

\begin{document}



\title{REINTERPRETING DARK ENERGY THROUGH BACKREACTION: THE MINIMALLY COUPLED MORPHON FIELD}

\author{JULIEN LARENA}

\address{Laboratoire de l'Univers et de ses th\'eoies (LUTH),\\
 CNRS UMR 8102, Observatoire de Paris and Universit\'e Paris 7 Denis Diderot\\
\email{julien.larena@obspm.fr}}

\author{THOMAS BUCHERT}

\address{Arnold Sommerfeld Center for Theoretical Physics ASC,\\
Ludwig--Maximilians--Universit\"{a}t, Theresienstra{\ss}e 37,\\
80333 M\"{u}nchen, Germany\\
\email{buchert@theorie.physik.uni-muenchen.de}}

\author{JEAN-MICHEL ALIMI}

\address{Laboratoire de l'Univers et de ses th\'eoies (LUTH),\\
 CNRS UMR 8102, Observatoire de Paris and Universit\'e Paris 7 Denis Diderot\\
\email{jean-michel.alimi@obspm.fr}}


\begin{abstract}
In the context of averaged cosmologies, the effective equations can be written in the form of "regional" Friedmannian equations with additional sources arising from the so-called backreaction of inhomogeneities. We propose a mean field description of this backreaction in terms of a regionally homogeneous scalar field: this provides a physical motivation to the phenomenological scalar fields generically called quintessence fields. We explicitly reconstruct the potential of the scalar field for a one-parameter family of scaling solutions to the backreaction problem, showing that it entails most of the standard scalar fields including e.g. standard and phantom quintessence scenarii.
\end{abstract}
\bodymatter

\section{Introduction}\label{intro}

Regionally averaged relativistic cosmologies are a way to tackle the so-called coincidence problem of standard Friedmannian cosmologies:  why is the expansion accelerating approximately at the same time when the Universe becomes structured, that is when the density contrast in the matter field is no longer small on a wide range of scales? 
The main issue is to try and link the dynamics of the Universe on large scales to its structuration on smaller scales\cite{backreactionsbiblio1}. It consists in building, from a fully non-homogeneous Universe, cosmological models that are homogeneous, thanks to a spatial averaging procedure. It results in equations for a volume scale factor that not only include an averaged matter source term, but also additional terms that arise from the coarse-grained inhomogeneities. These additional terms are commonly named backreaction.
In this contribution, we review a correspondence, proposed and discussed in a recent paper\cite{Buchert2006} between regionally averaged cosmologies and Friedmannian scalar field cosmologies, the scalar field being interpreted in this context as a mean field description of the inhomogeneous Universe, that can play the role of a field responsible for the dark energy phenomenon.
\section{Regionally averaged cosmologies: the backreaction context}
In this paper, since we are interested in the late time behavior of the
cosmological model, we restrict the analysis to a Universe filled with an
irrotational fluid of dust matter with density $\rho (t,X^{i})$.
Foliating the spacetime by flow-orthogonal hypersurfaces with the
3-metric $g_{ij}$ (that has no a priori symmetry), the line element reads $ds^{2}=-dt^{2}+g_{ij}dX^{i}dX^{j}$.
The large scale homogeneous model is then built by averaging the scalar parts of the general relativistic equations on a spatial domain $\CD$ with a spatial averager applied to any scalar function $\Upsilon (t,X ^{i})$: 
\begin{equation}
\label{averager}
\average{\Upsilon (t,X ^{i})}=\frac{1}{V_{\CD}}\int_{\CD}\Upsilon (t,X^{i})Jd^{3}X\mbox{ ,}
\end{equation}
 where $V_{\CD}$ is the volume of the domain $\CD$ and $J=\sqrt{\mbox{det}(g_{ij})}$.
 Then, one can define a volume scale factor $a_{\CD}=(V_{\CD}/V_{\initial\CD})^{1/3}$ \footnote{This volume scale factor has a simple physical interpretation, since the averaged dust density evolves as $\average{\rho}\propto a_{\CD}^{-3}$, due to our foliation that is locally comoving with the fluid.}, and one can build an effective homogeneous cosmological model on the domain $\CD$. The system of equations for that model characterizes the properties of the Universe on large scales corresponding to the scale of the domain $\CD$. It preserves the main feature of the standard FRW Universe: the properties of the Universe on large scales can be deduced from a single scale factor; but, due to the non-commutativity of the spatial averaging and the time derivatives,  this scale factor now obeys dynamical equations given by (\ref{averagedsystem}) that differ from the FRW equations for a dust fluid because of additional source terms, ${\cal Q}_{\CD}=2\average{(\theta-\average{\theta})^{2}}/3-2\average{\sigma^{2}}$, called kinematical backreaction, that are essentially the spatial variances on $\CD$ of the local expansion rate $\theta$ and of the rate of shear $\sigma$: 
\begin{eqnarray}
\label{averagedsystem}
\left(\frac{\dot{a}_{\CD}}{a_{\CD}}\right)^{2}&=&\frac{8\pi
  G}{3}\average{\rho}-\frac{\average{\CR}+{\cal Q}_{\CD}}{6}\nonumber\\
\frac{\ddot{a}_{\CD}}{a_{\CD}}&=&-\frac{4\pi G}{3}\average{\rho}+\frac{{\cal
    Q}_{\CD}}{3}\\
a_{\CD}^{-6}\partial_{t}\left(a_{\CD}^{6}{\cal
  Q}_{\CD}\right)&=&-a_{\CD}^{-2}\partial_{t}\left(a_{\CD}^{2}\average{\CR}\right)\nonumber
  \mbox{ .}
\end{eqnarray}
Moreover, the averaged 3-curvature $\average{\CR}$ is no longer a constant curvature term because it explicitly couples to these variances (see the third equation of system (\ref{averagedsystem}), that is simply an integrability condition and is therefore not an independent equation).
One immediately infers from the second equation of (\ref{averagedsystem}) that an effective acceleration is possible iff ${\cal Q}_{\CD}>4\pi G \average{\rho}$. 
\section{Correspondence with scalar field cosmologies}
It has been shown \cite{Buchert2006} that the system describing averaged cosmologies can be cast into the form of a Friedmannian dust cosmology in the presence of an additional standard minimally coupled scalar field $\Phi_{\CD}$ with a self-interaction potential $U_{\CD}(\Phi_{\CD})$, this scalar field representing the backreaction and averaged 3-curvature effects. The potential $U_{\CD}(\Phi_{\CD})$ is identical, up to a constant factor, with the averaged 3-curvature and the kinetic energy of the scalar field is a linear combination of the backreaction and averaged curvature. Thanks to that correspondence, one can infer the dynamical properties of the backreaction and averaged curvature by using all the constraints that are known on quintessence fields in the context of Friedmannian cosmologies. For example, if one knows the dark energy equation of state in a particular quintessence model, one can find the evolution of the backreaction that leads to such an effective equation of state.
For example particular solutions have been studied \cite{Buchert2006}: considering backreaction and averaged curvature that are power laws of the effective volume scale factor, the potential $U_{\CD}(\Phi_{\CD})$ has been reconstructed, and it corresponds to known quintessence models \cite{Staroetal}. Apart from the quasi-Friedmannian case in which the averaged curvature and the backreaction decouple, the ratio between backreaction and averaged curvature ($r_{\CD}={\cal Q}_{\CD}/\average{\CR}$) is constant, implying a strong coupling between them. Then, the properties of the corresponding scalar field essentially depends on this ratio, and it has been shown that varying this ratio can lead to a scale-dependent cosmological constant (for $r_{\CD}=-1/3$), and to almost all types of dark energy scalar fields such as standard quintessence (for $r_{\CD}\in ]-1/3,0[$) and phantom fields (for $r_{\CD}\in ]-1,-1/3[$). So, in this class of solutions, the measure of the dark energy equation of state today can fix the ratio between backreaction and averaged 3-curvature that is necessary today to explain dark energy.
Finally, it is interesting to note that in the phase plane of these scaling solutions, the Einstein-de Sitter scenario is a saddle point for the dynamics, so that a small amount of initial backreaction and/or averaged curvature naturally pushes the system far from it and may lead to a late-time accelerating phase. The open question is whether this mechanism can produce a quantitatively sufficient amount today, a subject of ongoing work \cite{buchert:brazil}.


\vfill


\begin{thebibliography}{00}

\bibitem{backreactionsbiblio1} G. F. R. Ellis, {\it General Relativity and Garvitation}, Dordrecht: Reidel (1984); T. Buchert, {\it Gen. Rel. Grav.} {\bf 32}, 105 (2000); T. Buchert, {\it Gen. Rel. Grav.} {\bf 33}, 1381 (2001), T. Buchert, {\it Class. Quant. Grav.} {\bf 23}, 817 (2006); E. W. Kolb, S. Matarrese and A. Riotto, {\it New J. Phys.} {\bf 8}, 322 (2006); T. Buchert, {\it Class. Quantum Grav.} {\bf 22}, L113 (2005); Y. Nambu and M. Tanimoto, {\it gr-qc/0507057} (2005); A. Paranjape and T.P. Singh, {\it astro-ph/0605195} (2006); S. R\"as\"anen, {\it Class. Quantum Grav.} {\bf 23}, 1823 (2006); S. R\"as\"anen, {\it J.C.A.P.} {\bf 0611}, 003 (2006); G. F. R. Ellis and T. Buchert, {\it Phys. Lett. A} {\bf 347}, 38 (2005); M.-N. C\'el\'erier, {\it astro-ph/0612222}, contribution to this volume
\bibitem{Buchert2006} T. Buchert, J. Larena and J.-M. Alimi, {\it Class. Quantum Grav.} {\bf 23}, 6379-640 (2006)

\bibitem{Staroetal} V. Sahni and A. A. Starobinskii, {\it Int. J. Mod. Phys. D} {\bf 9}, 373 (2000); V. Sahni, T. D., A. A. Starobinskii and U. Alam, {\it JETP Lett.} {\bf 77}, 201 (2003); L. A. Urena-Lopez and T. Matos, {\it Phys. Rev. D} {\bf 62}, 081302 (2000)

\bibitem{buchert:brazil} T. Buchert, {\it gr-qc/0612166}

\end{thebibliography}
\end{document}